\documentclass{emulateapj} 
\usepackage{apjfonts}
\usepackage{amsmath}
\begin{document}
\slugcomment{ApJ (Letters), in press}
\newcommand{\LeeZinn}{\mathcal{L}}
\newcommand{\msait}{MmSAIt}


\shortauthors{M. Catelan et al.} 
\shorttitle{Constraints on Helium Enhancement in M3}

\title{Constraints on Helium Enhancement in the Globular Cluster M3 (NGC~5272):\\
       The Horizontal Branch Test}

\author{M. Catelan\altaffilmark{1,2,3}}

\author{F. Grundahl\altaffilmark{4}}

\author{A. V. Sweigart\altaffilmark{5}}

\author{A. A. R. Valcarce\altaffilmark{2}}

\author{C. Cort\'es\altaffilmark{2,6}}

\altaffiltext{1}{John Simon Guggenheim Memorial Foundation Fellow} 

\altaffiltext{2}{Departamento de Astronom\'{i}a y Astrof\'{i}sica, Pontificia Universidad
Cat\'olica de Chile, Av. Vicu\~na Mackenna 4860, 782-0436 Macul, Santiago, Chile} 

\altaffiltext{3}{On sabbatical leave at Catholic University of America, 
Department of Physics, 200 Hannan Hall, Washington, DC 20064; and Michigan State University, 
Department of Physics and Astronomy, 3215 Biomedical and Physical Sciences Bldg., East 
Lansing, MI 48824} 

\altaffiltext{4}{Department of Physics \& Astronomy, Aarhus University, Ny Munkegade, 
8000 Aarhus C, Denmark} 

\altaffiltext{5}{NASA Goddard Space Flight Center, Exploration of the Universe Division, 
   Code 667, Greenbelt, MD 20771} 

\altaffiltext{6}{Departamento de F\'{i}sica Te\'orica e Experimental, Universidade Federal 
do Rio Grande do Norte, Campus Universit\'ario, 59072-970 Natal, RN, Brazil}

\begin{abstract}
It has recently been suggested that the presence of 
multiple populations showing various amounts 
of helium enhancement is the rule, rather than the exception,
among globular star clusters. 
An important prediction of this helium enhancement scenario is that 
the helium-enhanced blue horizontal branch (HB) stars should be brighter 
than the red HB stars which are not helium-enhanced. In this 
{\em Letter}, we test this prediction 
in the case of the Galactic globular cluster 
M3 (NGC~5272), for which the helium-enhancement scenario predicts 
helium enhancements of $\gtrsim 0.02$ in virtually all blue HB stars. 
Using high-precision \citeauthor{bs63} photometry and spectroscopic 
gravities for blue HB stars, we find that any helium enhancement among 
most of the cluster's blue HB stars is very likely less than 0.01, 
thus ruling out the much higher helium enhancements that have been proposed 
in the literature. 
\end{abstract}

\keywords{stars: abundances --- Hertzsprung-Russell diagram --- 
          stars: evolution --- stars: horizontal-branch --- 
		  (Galaxy:) globular clusters: general --- (Galaxy:) globular clusters: 
		  individual (M3~= NGC~5272, M13~= NGC~6205) 
          }

\section{Introduction}
Globular star clusters (GC's) have traditionally been assumed to be excellent 
approximations to so-called ``simple stellar populations,'' which are idealized 
systems in which all stars were formed at precisely the same time, from a 
chemically homogeneous cloud. However, recent observations, both photometric
and spectroscopic, have cast serious doubts on this long-standing paradigm. 

The presence of large 
abundance anomalies was first identified in the GC
$\omega$~Centauri (NGC~5139). In this cluster, not only such  
light elements as C, N, O, F, Na, Mg, and Al, but also the Fe-peak, 
s-process, and r-process elements, are seen to vary by large amounts 
from one star to the next \citep[e.g.,][ and the extensive list of 
references provided therein]{cjea08}. Many of these abundance patterns  
seem to extend all the way down to the main sequence \citep[e.g.,][]{lsea08}. 
This strongly suggests that such abundance anomalies owe their origin to 
multiple star formation episodes within the cluster, each accompanied by 
a corresponding enrichment of the intracluster medium by the ejecta of 
the massive and intermediate-mass stars which were formed in the prior
stellar generations. 
While a large spread in Fe-peak abundances has only been 
detected in $\omega$~Cen, many of $\omega$~Cen's abundance anomalies 
have also been found in other clusters, albeit often at a (much) less 
dramatic level (e.g., \citeauthor{csea04} \citeyear{csea04}; 
\citeauthor{cjea05} \citeyear{cjea05};
\citeauthor{dyea09} \citeyear{dyea09}; see also 
\citeauthor{rgea04} \citeyear{rgea04}, for a recent review). 

Analyses of the multiple main sequences found in the deep color-magnitude diagrams
(CMD's) of $\omega$~Cen strongly suggest that the {\em helium abundance} $Y$ in 
the cluster may also have changed dramatically from one star formation 
episode to the next \citep[e.g.,][]{jn04,fdea05,gpea05}. Recent evidence 
suggests that other clusters may also show sizeable $Y$ variations. According 
to the deep CMD analysis of NGC~2808 by \citet{gpea07}, multiple populations 
with different helium abundances are also present in this cluster. Strong 
arguments have also been raised in favor of helium enhancements among 
at least some of the stars in NGC~6388 and NGC~6441 \citep[see, e.g.,][ and 
references therein]{mcea06,cd07}. Interestingly, these four GC's figure among 
the most massive of all Galactic GC's. 

Very recently, it has been suggested that multiple star formation episodes in GC's, 
accompanied by widely different amounts of helium enrichment, are in fact not the 
exception, but instead the rule \citep{dac08}. In the particular 
case of M3 (NGC~5272), such a claim had previously been made also 
by \citet{cd08}, on the basis of an analysis of the period distribution of M3's 
RR Lyrae variables and the color extension of the HB blueward of the 
instability strip. Due to the recognized need to assume a very sharply 
peaked mass distribution to account for the sharply peaked shape of the RR Lyrae 
period distribution in M3 \citep{rc89,mc04,mcea05}, 
\citeauthor{dac08} suggested that 
the mass distribution of HB stars is {\em always} sharply peaked, and that 
the color spread routinely observed in GC CMD's is instead due to internal 
variations in the helium abundance $Y$. It is important to emphasize that 
such a spread in $Y$ is {\em not} required to explain the observed 
period distributions. Rather, the case for helium enhancement, at least 
in the case of M3-like clusters, rests almost entirely on the presence of 
color spreads among HB stars.

\begin{figure*}[ht]
  \centerline{
  \includegraphics*[width=6.825in]{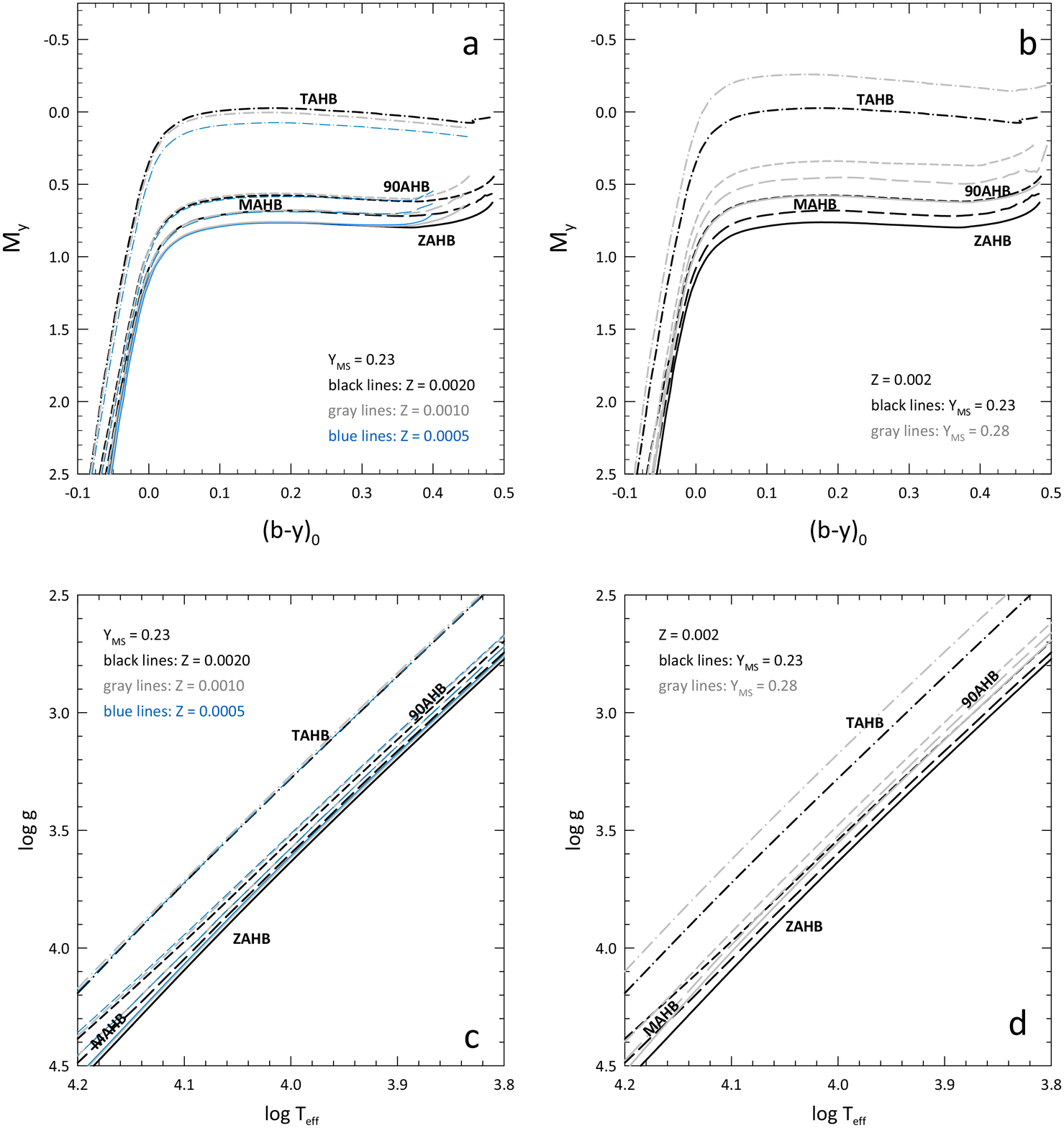}
  }
  \caption{{\em Panel a}: comparison between fiducial HB loci (ZAHB, 
    MAHB, 90AHB, TAHB; see definitions in \S\ref{sec:models}) for three 
    different metallicities ({\em black lines}: $Z = 0.002$; {\em gray
	lines}: $Z = 0.001$; {\em blue lines}: $Z = 0.0005$), in the 
	$M_y$, $(b\!-\!y)_0$ diagram. To 
	produce this plot, the $Z = 0.001$ loci were shifted in $M_y$ by 
	$+0.075$~mag, and the $Z = 0.0005$ loci by $+0.140$~mag. 
    Note that with these shifts all three of the ZAHB's coincide, 
	except at the extreme red end of the HB, where no unevolved 
	stars are found in M3. 
	{\em Panel b}: as in panel a, but for a fixed metallicity and two 
	different helium abundances ({\em black lines}: $Y = 0.23$; 
	{\em gray lines}: $Y = 0.28$), and without applying any 
	shifts to the CMD positions. 
	Panels {\em c} and {\em d}: as in a and b, but in the 
	$\log g$, $\log T_{\rm eff}$ plane, and without applying any shifts
	to the data. 
   }
      \label{fig:ztest}
\end{figure*}

\begin{figure}[ht]
  \centerline{
  \includegraphics[width=2.84in]{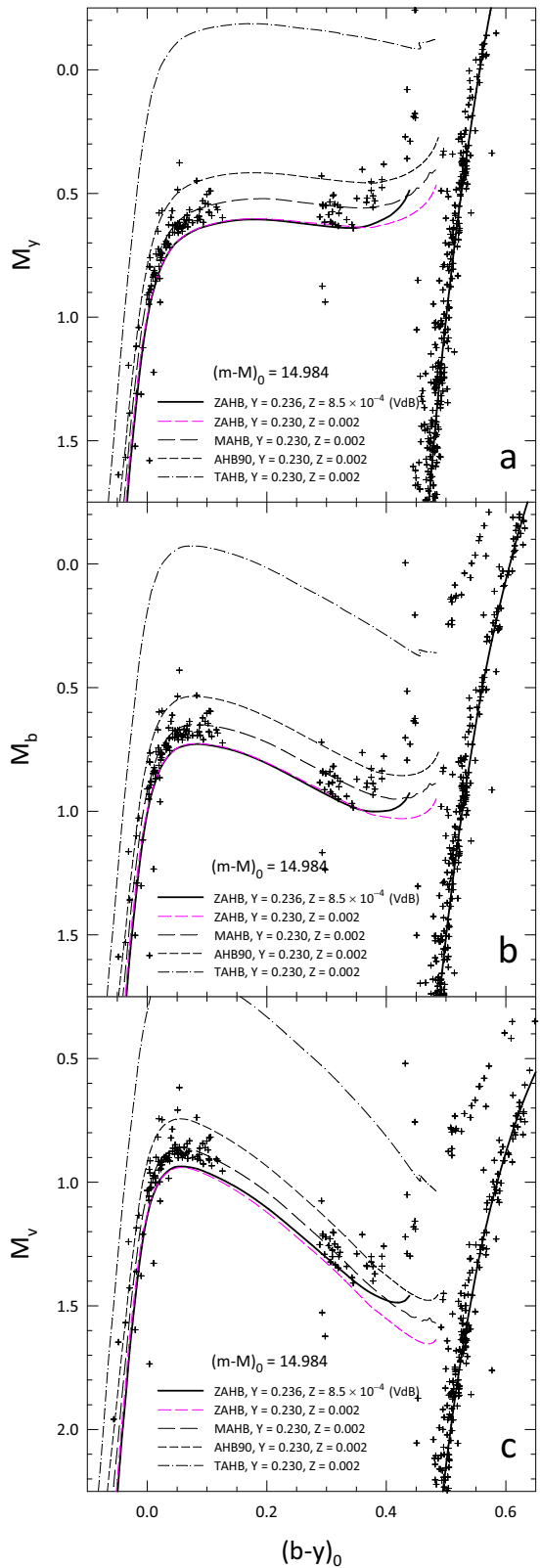}
  }
  \caption{Comparison between fiducial HB sequences (see text) 
    and the empirical data for M3, in the 
	$M_y$, $(b\!-\!y)_0$ plane (panel {\em a}),
	$M_b$, $(b\!-\!y)_0$ plane (panel {\em b}), and 
	$M_v$, $(b\!-\!y)_0$ plane (panel {\em c}). 
	The empirical data were shifted vertically so as to 
	produce a satisfactory match to the theoretical red ZAHB, as
	given by the \citet{vdbea06} models for 
	${\rm [Fe/H]} = -1.61$, 
	$[\alpha/{\rm Fe}] = +0.3$, and 
	$Y = 0.236$,
	in 
	the $M_y$, $(b\!-\!y)_0$ plane. The \citeauthor{vdbea06} RGB
    sequence is also displayed.
   }
      \label{fig:singley}
\end{figure}

\begin{figure*}[ht]
  \centerline{
  \includegraphics*[width=6.5in]{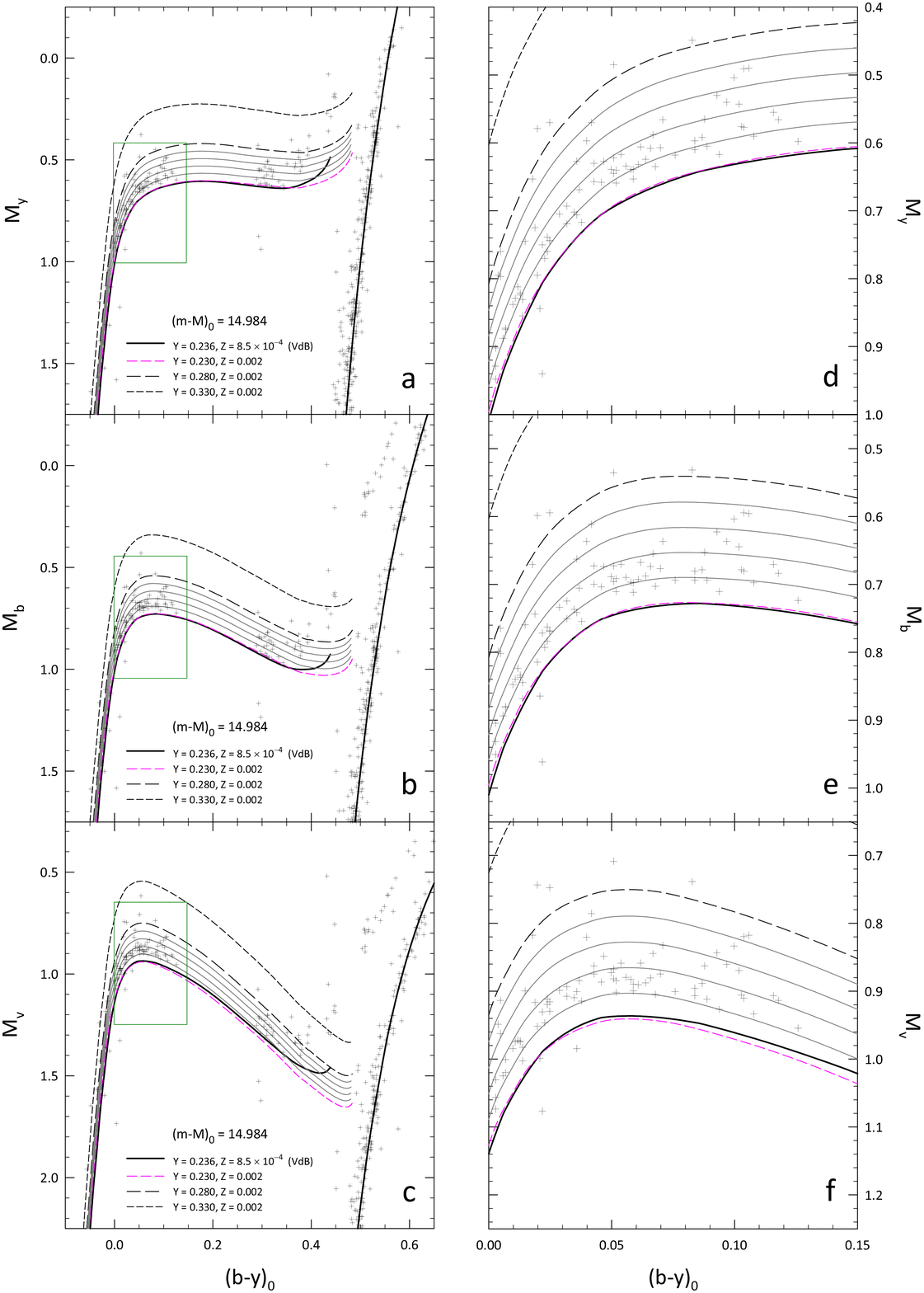}
  }
  \caption{
    As in Figure~\ref{fig:singley}, but plotting only the 
    ZAHB's, for the several different $Y$ values as indicated. 
	The {\em thin solid lines} indicate interpolated ZAHB loci 
	for $Y$ values between 0.23 and 0.28, in intervals 
	of 0.01. Panels {\em d}~through {\em f} present 
	expanded views of the rectangular box around the blue HB 
	``knee'' in panels~{\em a} through {\em c}, respectively.
   }
      \label{fig:manyy}
\end{figure*}

If confirmed, this scenario would not only have major implications for our 
understanding of how GC's form, but would also represent 
a major shift from the canonical paradigm, dominant 
since the late-1960's/early-1970's \citep{vcea69,ir70,jf72,rtr73}, which  
ascribes the color spread seen among HB stars to the stochastic 
nature of mass loss on the red giant branch (RGB). In the canonical scenario, 
blue HB stars lose significantly more mass as they climb up the RGB than do their 
red HB counterparts. In contrast, in the \citet{dac08} scenario blue HB 
stars, even at the ``horizontal'' level of the HB, owe their blue colors to a 
higher initial $Y$. In the specific case of M3, the predicted He enhancement,  
even in the ``horizontal'' blue HB stars, should fall in the range between 
0.02 and 0.025, according to Fig.~4 of \citet{cd08}. 

Fortunately, the $Y$ enhancement scenario can be tested, using a variety of 
photometric and spectroscopic tools, especially for these moderately cool 
blue HB stars at the ``horizontal'' level. 
In particular, it is well known that HB stars 
with higher $Y$ are brighter, at a given $T_{\rm eff}$, due to their stronger 
H-burning shells \citep[e.g.,][]{sg76}. Therefore, in the \citet{dac08} 
scenario bluer HB stars should be brighter than their redder counterparts. 
While this may help account for the sloping natures of the HB's in 
the moderately metal-rich GC's NGC~6388 and NGC~6441
\citep[e.g.,][ and references therein]{gbea07}, 
such a CMD test has never being carried out for 
most of the GC's that, according to \citeauthor{dac08}, present multiple 
He-enhanced populations. 

The purpose of this {\em Letter} is 
accordingly to present a first test of the helium enhancement scenario, in the 
specific case of M3, by comparing high-precision photometry for the cluster 
in the \citet{bs63} system, as well as spectroscopically derived gravities, 
with theoretical HB models computed for a variety of $Y$ values.

\section{Observational Data}
The CMD data used in this paper were taken from \citet{fgea98,fgea99}, 
to which the reader is referred for additional details regarding the 
calibration of our photometry. Briefly, our M3 photometry is based 
on a series of images obtained on the Nordic Optical Telescope, 
using the thinned AR-coated $2048\times 2048$ pixel HiRAC CCD camera, 
with $0.11\arcsec$ pixel size, thus covering
approximately $3.75\arcmin$ on a side. Two overlapping fields were
observed, with one field centered on the cluster center to ensure a 
large sample of HB and RGB stars. 

In addition to the CMD data, we also used the gravities and temperatures 
derived by \citet{bb03}, on the basis of observations 
using the HIRES cross-dispersed echelle spectrograph on the Keck~I 
telescope.

\section{Theoretical Models}\label{sec:models}
In the present paper, we use the evolutionary tracks computed by \citet{mcea98} 
and \citet{sc98} for scaled-solar, heavy-element abundances $Z$ of 
$0.0005$, $0.001$, and $0.002$. Evolutionary tracks for a He abundance 
$Y = 0.23$ by mass were computed for each of these $Z$ values. 
Additional He-enhanced tracks for $Y = 0.28$, 0.33 were also 
computed for $Z=0.002$.\footnote{Note that these helium abundances 
refer to the initial main sequence values, and thus do not include 
the small increase that occurs during the dredge-up phase on the RGB.} 
The theoretical models were transformed to the 
\citet{bs63} $uvby$ system by using the color transformations and 
bolometric correction tables provided by \citet{jcea04}. The same procedures 
were also adopted in the recent work by \citet{cac08} and \citet{cc08}, where
the period-color and period-luminosity relations of RR Lyrae stars in the 
\citeauthor{bs63} system were presented. In addition, we also use a set 
of models computed by \citet{vdbea06}, for a chemical composition 
${\rm [Fe/H]} = -1.614$, $[\alpha/{\rm Fe}] = +0.3$ (which translates, 
according to those authors, to a $Z = 8.45\times 10^{-4}$), and 
$Y = 0.236$.   

M3 has a metallicity of ${\rm [Fe/H]} = -1.57$ in the \citet{zw84} scale, 
and ${\rm [Fe/H]} = -1.34$ in the \citet{cg97} scale. Taking into account 
an enhancement of the $\alpha$-capture elements by 
$[\alpha/{\rm Fe}] = +0.27$ \citep{bc96}, and using the relation between 
$Z$, [Fe/H], and $[\alpha/{\rm Fe}]$ from \citet{msea93}, we find  
an overall metallicity of 
$Z = 8.3 \times 10^{-3}$ and 
$Z = 1.4 \times 10^{-3}$, respectively, on these two metallicity scales. 
Therefore, our  models 
comfortably bracket the range of possible metallicity values for the cluster. 

Since we are primarily interested in constraining the {\em change} in 
$Y$ between the red and blue HB, the exact choice of $Z$ value 
is basically irrelevant for our purposes. This is shown in 
Figure~\ref{fig:ztest}{\em a}, 
where our models for $Z = 0.0005$, 
$Z = 0.001$, and $Z = 0.002$ are compared in 
the $M_y$, $(b\!-\!y)_0$ diagram. To produce this plot, we 
first registered the low-metallicity zero-age HB (ZAHB) sequences 
to the $Z = 0.002$ ZAHB by shifting the $Z = 0.0005$ ZAHB 
by $+0.140$~mag and the $Z = 0.001$ ZAHB by $+0.075$~mag in $M_y$. 
We then applied the same shifts to the evolved HB sequences for each 
$Z$ value, and derived several fiducial loci representing evolved HB 
stars, as follows: i)~MAHB, standing for the {\em middle-age HB}, or HB ridgeline, 
which gives the average position occupied by all HB stars, assuming a uniform mass 
distribution along the entire ZAHB; ii)~90AHB, or {\em 90\%-age HB}, which is 
approximately the locus below which one should find $\approx 90\%$ of all HB 
stars;\footnote{In like vein, one may define the ``50\%-age HB'' locus, or
50AHB, as the locus occupied by HB stars of different masses which have 
completed 50\% of their HB evolution. Note, however, that this is not
exactly coincident with the MAHB; since the distribution of HB luminosities
is not Gaussian, the MAHB is in fact slightly more luminous than 
the corresponding 50AHB.} iii)~TAHB, or 
{\em terminal-age HB}, which is simply the He exhaustion locus.  

This registration procedure clearly leads to an excellent match over 
the entire range of ZAHB colors and HB luminosities, except (as expected) 
at the extreme red end of the HB and close to the TAHB. Since M3 
lacks unevolved HB stars at colors redder than about $(b\!-\!y)_0 \simeq 0.34$, 
and since the TAHB lies very far away from where most HB stars are found 
(see Fig.~\ref{fig:singley}), these differences are of no concern 
for our purposes. Similar results follow when using $M_b$ and $M_v$, but 
not $M_u$, which we have found to be affected by strong metallicity 
effects. Since the $v$- and (especially) the $u$- band bolometric 
corrections from \citet{jcea04} may be quite uncertain (D. A. VandenBerg, 
priv. comm.), we have decided not to include our $u$-band photometry in this 
{\em Letter} \citep[but see][]{mc09}. We will, however, include our $v$-band 
data, but caution that the comparison of these data with the models is less 
reliable than for the $y$ or $b$ data.

While the metallicity effects are clearly mild, both the ZAHB and 
the evolved HB loci depend strongly on $Y$, as can be easily inferred 
from Figure~\ref{fig:ztest}{\em b}. This shows that a {\em differential} 
comparison between the luminosities of red and blue HB stars, around the 
metallicity of M3, should provide a strong indicator of any possible 
helium enhancement, with only a very mild dependence on metallicity. 

Panels~{\em c} and {\em d} in Figure~\ref{fig:ztest} reveal the effects of 
metallicity and $Y$, respectively, on the $\log g - \log T_{\rm eff}$ plane. 
This again shows that metallicity plays but a minor role in defining
the position of a star on this plane, compared with $Y$.

\begin{figure}[ht]
  \centerline{
  \includegraphics[width=3.375in]{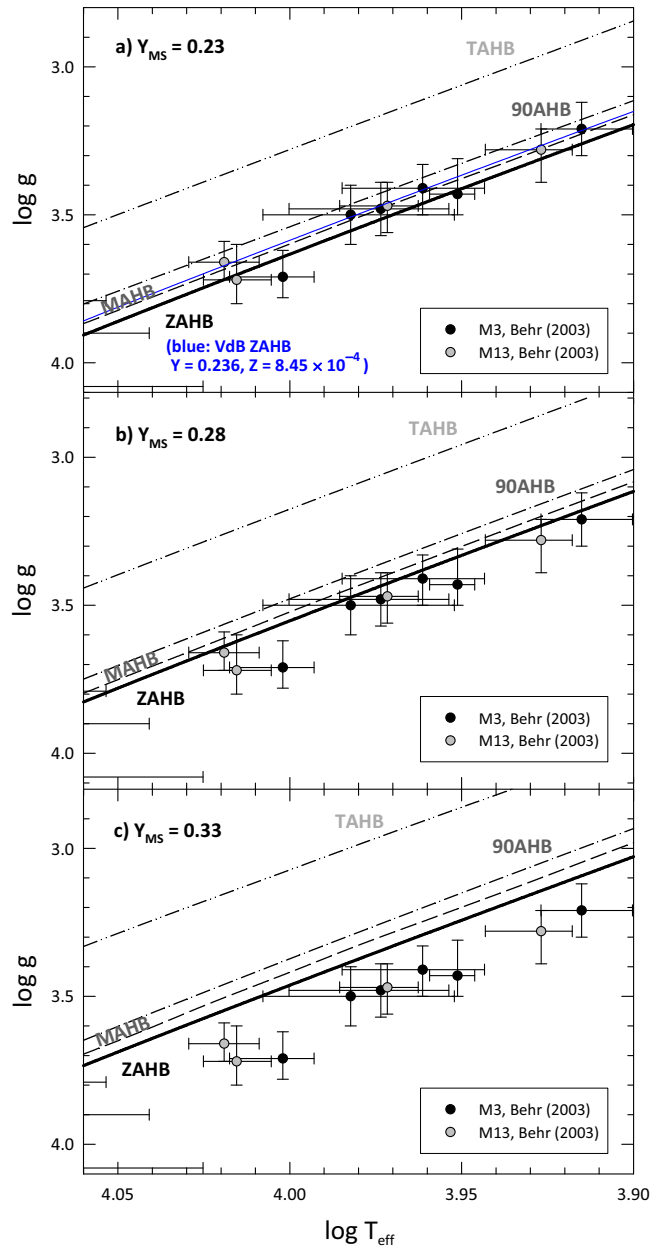}
  }
  \caption{Comparison between predicted and observed loci in the 
    $\log g$, $\log T_{\rm eff}$ plane, for a metallicity 
	$Z = 0.002$ and three different helium abundance values: 
	0.23 (panel {\em a}), 	
	0.28 (panel {\em b}), and	
	0.33 (panel {\em c}). Empirical data are provided for M3
    ({\em black circles}) and M13 ({\em gray circles}), from 
    \citet{bb03}. The blue line in panel {\em a} shows the ZAHB locus 
	of \citet{vdbea06} models for a $Y = 0.236$, 
    ${\rm [Fe/H]} = -1.61$, $[\alpha/{\rm Fe}] = +0.3$. 	
   }
      \label{fig:loggtest}
\end{figure}

\section{Results}\label{sec:results}
We compare the model predictions for fixed (low) $Y$ with the empirical CMD 
data in Figure~\ref{fig:singley}. To produce these plots, we have first corrected
the empirical data for reddening using a $E(\bv) = 0.01$, taken from \citet{wh96},  
and the extinction coefficients for the \citeauthor{bs63} system summarized 
in \citet{cac08}. We then shifted the data vertically, as required to match
the theoretical red ZAHB computed by \citet{vdbea06} to the lower envelope 
of the observed distribution in the $M_y$, $(b\!-\!y)_0$ plane. Note that these
\citeauthor{vdbea06} evolutionary tracks also provide an excellent fit to the 
RGB of the cluster. This results in a 
distance modulus of $(m\!-\!M)_0 = 14.984$ for M3. We then applied a shift 
of $\Delta {\rm mag} = -0.16$ to our $Z = 0.002$ models, in order
that they would match the \citeauthor{vdbea06} ones. 
As can be seen, such a procedure 
leads to an excellent morphological match between the \citeauthor{vdbea06} 
ZAHB for $Z = 8.45 \times 10^{-4}$ and our own for $Z = 0.002$, especially 
in the $M_y$, $(b\!-\!y)_0$ and $M_b$, $(b\!-\!y)_0$ planes, except again at
the very end of the red HB distribution, where no unevolved HB stars are found 
in M3. 

Except perhaps for a small deviation of the lower envelope of the blue 
HB stars in the immediate vicinity of the ``knee'' from the theoretical 
(single-$Y$) ZAHB, 
this figure shows a remarkable overall agreement between the model predictions
and the observations, without an immediately obvious 
need to invoke an increase in $Y$ for the blue HB stars. In addition, 
all the other HB fiducials for a fixed $Y$, including the MAHB, 90AHB, and 
TAHB, appear to describe the empirical data fairly well, without significant 
evidence for an excessive number of overluminous blue HB stars, as would be
expected in the helium-enhancement scenario (see Fig.~\ref{fig:ztest}{\em b}). 
To be sure, the evolutionary 
lifetimes along the canonical tracks do not provide a perfect match to the 
CMD distribution \citep{vc08}; however, it appears extremely unlikely 
that any such disagreements may somehow be due to internal variations in $Y$, 
since they seem to affect blue {\em and red} HB stars alike, as also noted by 
\citeauthor{vc08}. 

In Figure~\ref{fig:manyy} we compare our CMD data to ZAHB sequences for 
different $Y$ values in order to obtain a more quantitative limit on any 
increase in $Y$ between the red and blue HB populations. These ZAHB 
sequences are all for a fixed $Z = 0.002$ and have been shifted by 
$\Delta {\rm mag} = -0.16$ to match the \citet{vdbea06} ZAHB for 
$Z = 8.45 \times 10^{-4}$, as noted previously. 

Clearly, an enhancement in $Y$ by more than 0.01 in the blue HB stars 
would hardly be compatible with the data. In particular, we find many 
more stars below the blue ZAHB for $Y = 0.24$ than below the red ZAHB for 
$Y = 0.23$, 
again with the possible exception of the (more uncertain) 
$M_v$, $(b\!-\!y)_0$ plane.  
This strongly suggests that the level of He enhancement is most likely less 
than 0.01 among the cool blue HB stars in M3. 

Recall, from Figure~4 in 
\citet{cd08} and Table~1 in \citet{dac08}, that the bulk of the blue HB stars 
in M3 are predicted, in the helium-enhancement scenario, to be enriched in the 
range between 0.02 to 0.025. Such a level of helium enhancement is clearly 
ruled out by our data. Note also that, in the helium enhancement scenario, one 
should expect to see an increase in $Y$ towards bluer colors, but this is not 
supported by our data. 

Finally, Figure~\ref{fig:loggtest} compares the empirical and predicted 
positions of blue HB stars in M3 in the $\log g$, 
$\log T_{\rm eff}$ plane, for three different helium abundances, ranging 
from 0.23 (panel {\em a}) to 0.33 (panel {\em c}). To produce 
this plot, 
we restrict ourselves to temperatures cooler than 11,500~K, to avoid the 
well-known complications due to radiative levitation for hotter HB stars 
\citep{fgea99,gmea08}. Our CMD's indicate that less than 5\% of all of the 
HB stars in M3 are hotter than this limit, 
which corresponds to a color $(b\!-\!y)_0 \approx -0.025$. 

Similarly to what was found in our CMD analysis, the empirical gravities 
also seem entirely 
consistent with a uniform value of $Y$ among the blue HB stars in M3. 
In addition, there is no indication of an increase in $Y$ beyond the 
canonical value, a conclusion which becomes even stronger when the data 
are compared with the $\alpha$-enhanced ZAHB by \citet{vdbea06} for a 
$Y = 0.236$, $Z = 8.45 \times 10^{-4}$. 
Interestingly, the available data 
also suggest that at least the redder blue HB stars in M13 (NGC~6205) have a 
similar helium abundance as in M3. 
The sample size remains relatively small though, and therefore 
data for more stars in both clusters would certainly prove of interest 
to derive more conclusive results.

\section{Conclusions}
In this {\em Letter}, we have shown that high-precision, well-populated 
empirical CMD's, along with spectroscopic gravities, can be used to pose 
strong constraints on the presence of helium-enriched populations in GC's. 
Our results strongly suggest that any populations that may have formed after
a main initial burst in M3 likely preserved closely the same helium content 
as in the cluster's primordial gas, with a level of helium enhancement 
most likely not higher than 0.01.  
In future papers, we plan to apply similar tests to several other GC's.

\acknowledgments We thank an anonymous referee for some helpful comments,   
    and D. A. VandenBerg for some enlightening discussions. 
	Support for M.C. is provided
    by Proyecto Basal PFB-06/2007, by FONDAP Centro de Astrof\'{i}sica 15010003, 
	by Proyecto FONDECYT Regular \#1071002, and by a John Simon Guggenheim Memorial 
	Foundation Fellowship.

\end{document}